\documentclass{article}
\usepackage{graphicx}
\usepackage[english]{babel}
 
\usepackage{amsmath,amstext,amssymb}
 
\newcommand{\HOME}{/users/staf/wes/}
\newcommand{\lthreepaper}{\HOME/l3/paper/}
\newcommand{\lthreebiblio}{\lthreepaper biblio/}
\usepackage{cite}
\newcommand{\taumodel}{$\tau$-model}
\newcommand{\adhoc}{\textit{ad hoc}}

\newcommand{\ie}{{\it i.e.}}%
\newcommand{\vs}{{\it vs.}}%
\newcommand{\Eq}[1]{Eq.\,(\ref{#1})}%
\newcommand{\Eqs}[1]{Eqs.\,(\ref{#1})}%
\newcommand{\Fig}[1]{Fig.\,\ref{#1}}%
\newcommand{\Figs}[1]{Figs.\,\ref{#1}}%
\newcommand{\Pep}{e$^+$}%
\newcommand{\Pem}{e$^-$}%
\newcommand{\PZ}{\ensuremath{\mathrm{Z}}}
\newcommand{\eV}{\hbox{\ensuremath{\mathrm{e\kern-0.1em V}}}}%
\newcommand{\GeV}{\hbox{\ensuremath{\mathrm{G}}\eV}}%

\newcommand{\ycut}{\ensuremath{y_\mathrm{cut}}}
\newcommand{\ytt}{\ensuremath{y_{23}}}
\newcommand{\yttJ}{\ensuremath{y_{23}^\mathrm{J}}}
\newcommand{\pt}{\ensuremath{p_\mathrm{t}}}
\newcommand{\mt}{\ensuremath{m_\mathrm{t}}}

\newcommand{\LEP}{{\scshape lep}}
\newcommand{\ALEPH}{{\scshape aleph}}
\newcommand{\OPAL}{{\scshape opal}}
\newcommand{\DELPHI}{{\scshape delphi}}

\newcommand{\Lthree}{{\scshape l}{\small 3}}

\graphicspath{{.}{figs/}}       
 
\title{
\includegraphics[width=0.35\textwidth]{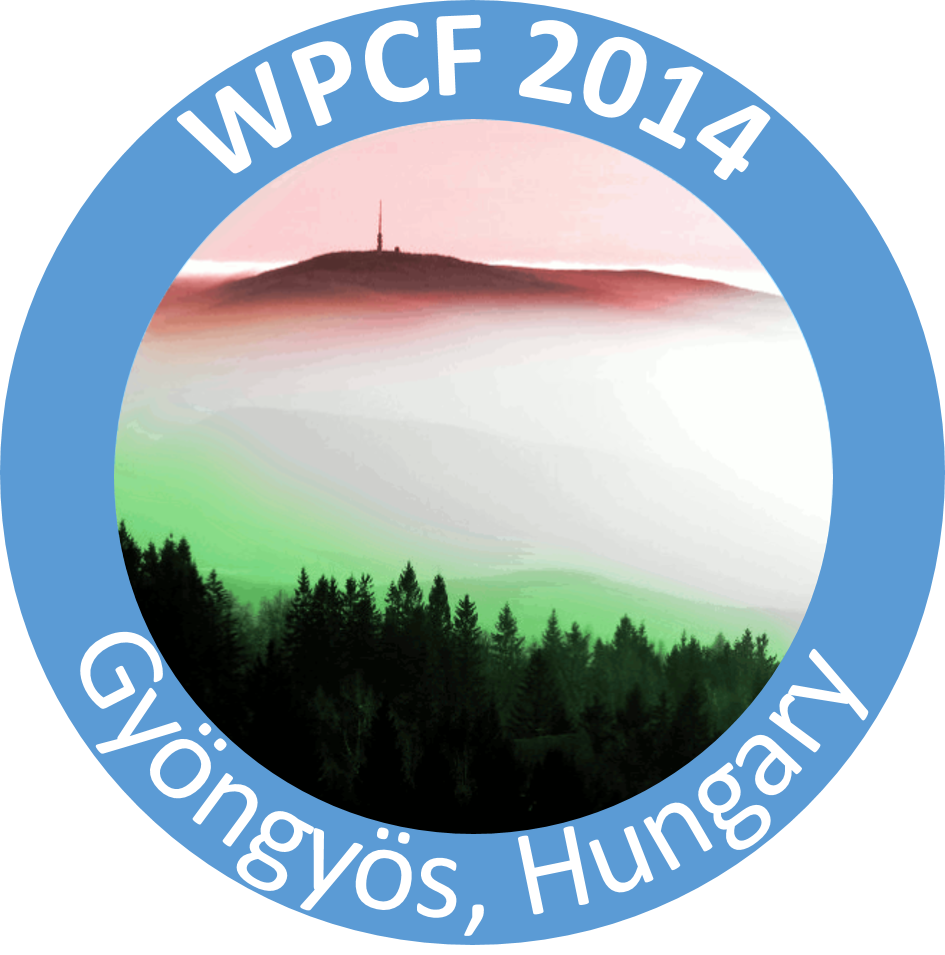}\\[1cm]
Multiplicity, Jet, and Transverse Mass dependence of Bose-Einstein Correlations in e$^{+}$e$^{-}$- Annihilation%
\footnote{This talk was also given at \textit{XLIV International Symposium on Multiparticle Dynamics}, Bologna, 8--12 September 2014.}}
\author{{Wesley J. Metzger$^1$,}\\[1ex]
$^1${IMAPP, Radboud University, 6525 AJ\ \ Nijmegen, The Netherlands}
}
 
\begin{document}
 
\fontfamily{lmss}\selectfont
\maketitle
 
\begin{abstract}
   Bose-Einstein correlations of pairs of identical charged pions
   produced in hadronic Z decays are analyzed for
   both two- and three-jet events.
   A parametrization suggested by the \taumodel\ is used to investigate the dependence of
   the Bose-Einstein correlation function on track multiplicity, number of jets, and transverse momentum.
\end{abstract}

\section{Introduction} \label{intro}
After a brief review of relevant previous results,
new \textit{preliminary\/} results are presented on the dependence of
the Bose-Einstein correlation function on track and jet multiplicity and transverse momentum,
using a parametrization
which has been found~\cite{L3_levy:2011} to describe well
Bose-Einstein correlations (BEC) in hadronic \PZ\ decay, namely that of
the \taumodel~\cite{Tamas;Zimanji:1990,ourTauModel},
 
\subsection{Review}    \label{review}
The    Bose-Einstein correlation function, $R_2$, is usually parametrized as
\begin{equation} \label{gauss_param}
  R_2 =    \gamma \left[ 1+ \lambda \exp \left(-\left(rQ\right)^{2} \right) \right] (1+ \epsilon Q) \;,
\end{equation}
and is measured by $R_2(Q)=\rho(Q)/\rho_0(Q)$,
where $\rho(Q)$ is the density of identical boson pairs with invariant four-momentum difference
$Q=\sqrt{-(p_1-p_2)^2}$
and $\rho_0(Q)$ is the similar density in an artificially constructed reference sample,
which should differ from the data only in that it does not contain BEC.
 
\begin{figure}      \centering
  \includegraphics*[width=.77\textwidth,angle=270,viewport=60 94 500 710,clip=true,scale=0.8]{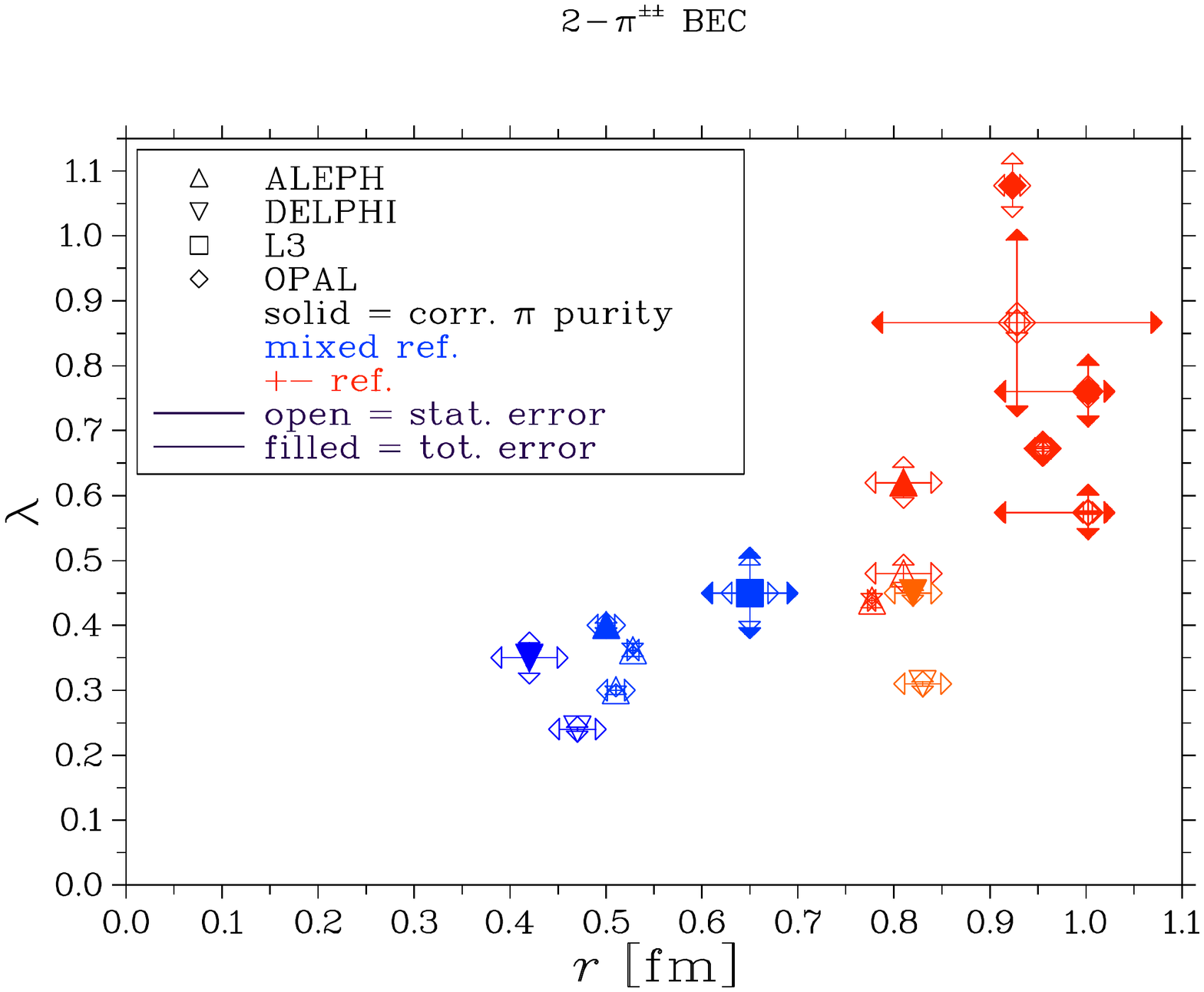}
  \caption{$\lambda$ and $r$ at $\sqrt{s}=M_\mathrm{Z}$ found in the \LEP\ experiments
             \cite{ALEPH:1992,ALEPH:2004,DELPHI:1992,L3_3pi:2002,OPAL:1991,OPALmult:1996,OPAL3D:2000}.}
  \label{fig:lrlep}
\end{figure}
\paragraph{Dependence on the reference sample}
Two methods were frequently used at \LEP\ to construct $\rho_0$:
unlike-sign pion pairs from the same event, and like-sign pairs from different events.
The latter method is generally referred to as mixed events.
However, it must be pointed out that the observed values of the parameters $r$ and $\lambda$
depend to a great extent on which reference sample is used.  This is clearly seen in
\Fig{fig:lrlep} where the values
of $\lambda$ and $r$ found for charged-pion pairs from hadronic \PZ\ decays
by the \LEP\ experiments
\ALEPH\ \cite{ALEPH:1992,ALEPH:2004},
\DELPHI\ \cite{DELPHI:1992}, \Lthree\ \cite{L3_3pi:2002} and
\OPAL\ \cite{OPAL:1991,OPALmult:1996,OPAL3D:2000} are displayed.
Solid points are corrected for pion purity; open points are not.
This correction increases the value of $\lambda$ but has little effect on the value of $r$.
All of the results with $r>0.7$ fm were obtained using an unlike-sign reference sample,
while those with smaller $r$ were obtained with a mixed reference sample.
The choice of reference sample clearly has a large effect on the observed values of $\lambda$ and~$r$.
In comparing results we must therefore be sure that the reference samples used
are comparable.
 
\begin{figure}      \centering
  \includegraphics*[width=.77\textwidth,angle=270,viewport=60 94 500 710,clip=true,scale=0.8]{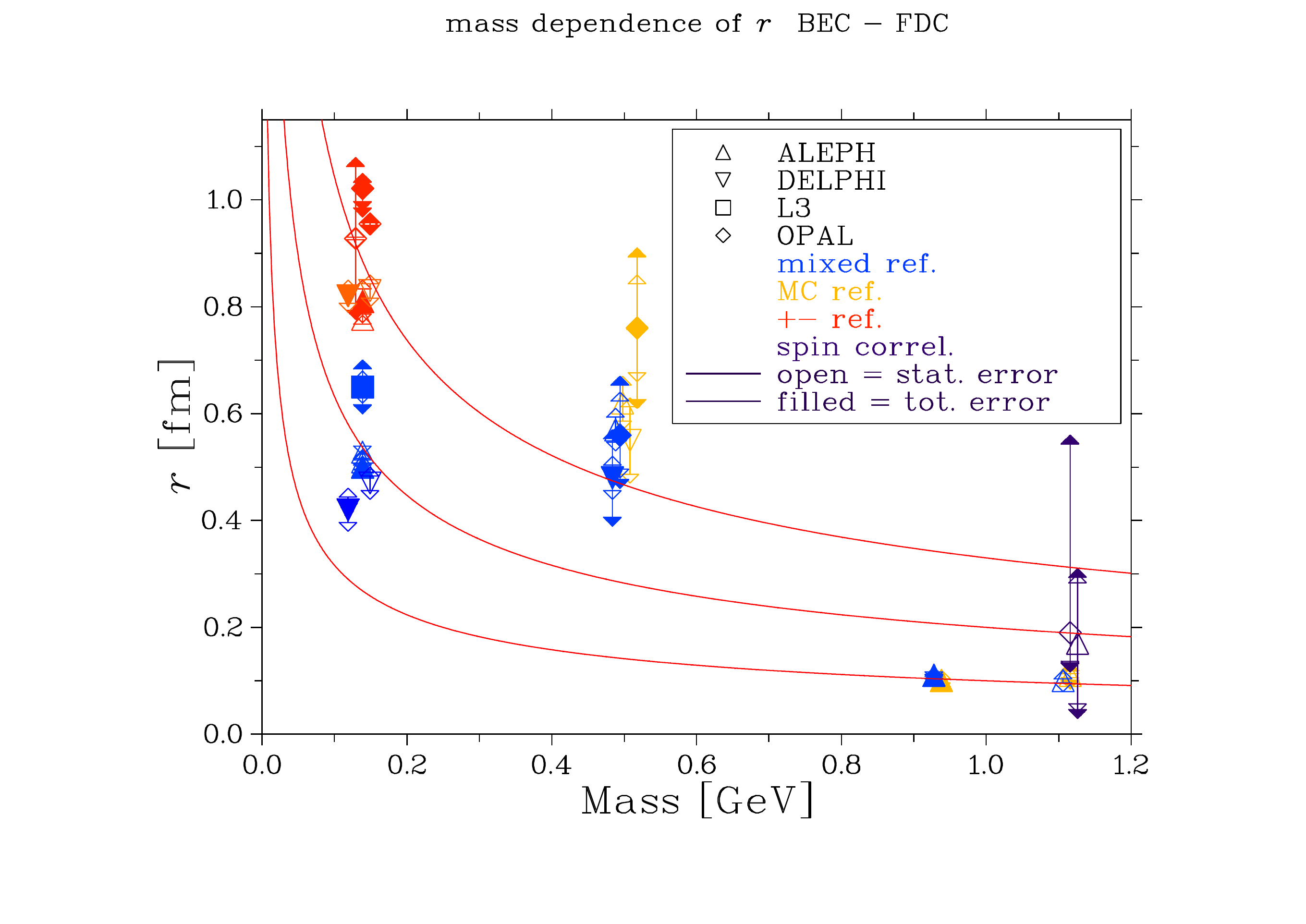}
  \caption{Dependence of $r$ on the mass of the particle as determined at $\sqrt{s}=M_\mathrm{Z}$ from
           2-particle BEC for charged pions
           \cite{ALEPH:1992,ALEPH:2004,DELPHI:1992,L3_3pi:2002,OPAL:1991,OPALmult:1996,OPAL3D:2000},
           charged kaons \cite{DELPHIK:1996,OPALKp:2001} and
           neutral kaons \cite{ALEPHnon_pi:2005,DELPHIK:1996,OPALK0:1995}
           and from Fermi-Dirac correlations for protons \cite{ALEPHnon_pi:2005}
           and lambdas \cite{ALEPHlam:2000,OPALlam:1996}.
           The curves illustrate a $1/\sqrt{m}$ dependence.
          }
  \label{fig:rmass}
\end{figure}
\paragraph{Dependence on the particle mass}
It has been suggested, on several grounds \cite{Alexander:2003}, that $r$ should depend on the particle mass
as $r\propto1/\sqrt{m}$.
Values of $r$ found at \LEP\ for various types of particle
are shown in Fig.~\ref{fig:rmass}.
Comparing only results using the same type of
reference sample (in this case mixed), we see no evidence for a $1/\sqrt{m}$ dependence.  Rather, the data
suggest one value of $r$ for mesons and a smaller value for baryons.  The value for baryons, about 0.1 fm,
seems very small, since the size of a proton is an order of magnitude greater.
If true it is telling us something unexpected about the mechanism of baryon production.
 
\begin{figure*}
\begin{minipage}{.323\textwidth}
 \centerline{{longitudinal}} 
 \includegraphics[width=0.77\textwidth,angle=270,viewport=80 120 521 715,clip]{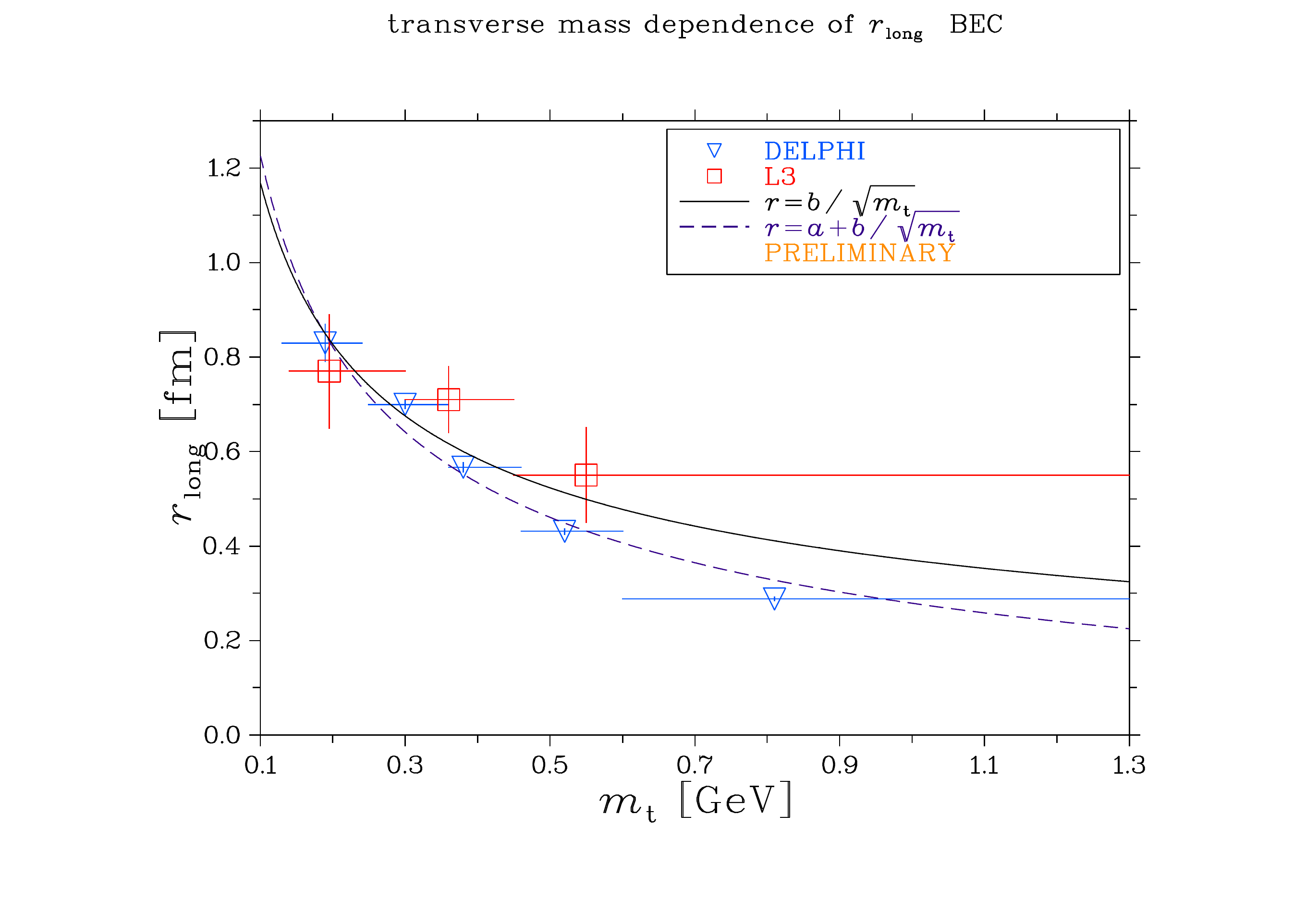}
\end{minipage}
\begin{minipage}{.323\textwidth}
 \centerline{{side}} 
 \includegraphics[width=0.77\textwidth,angle=270,viewport=80 120 521 715,clip]{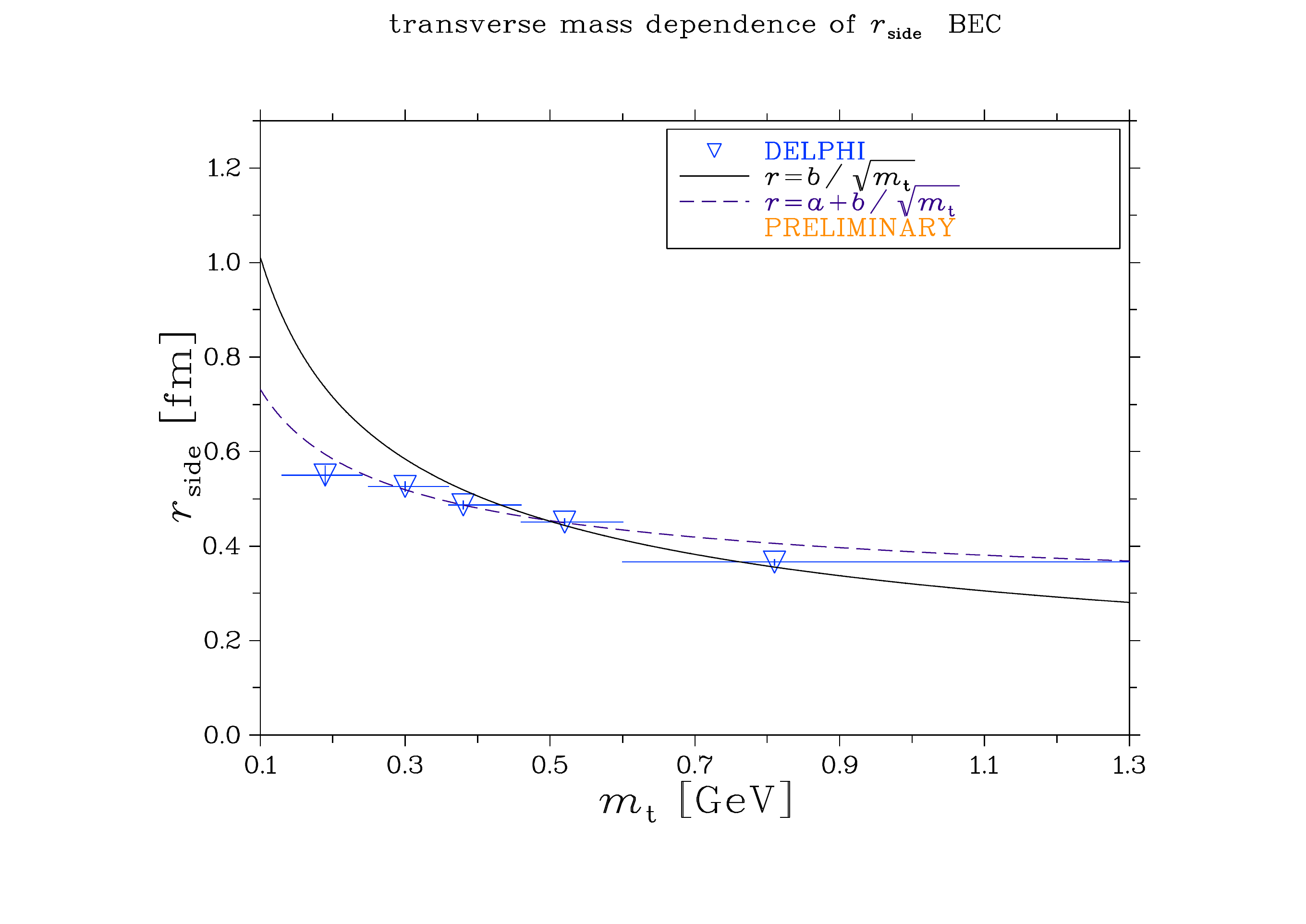}
\end{minipage}
\begin{minipage}{.323\textwidth}
 \centerline{{out}} 
 \includegraphics[width=0.77\textwidth,angle=270,viewport=80 120 521 715,clip]{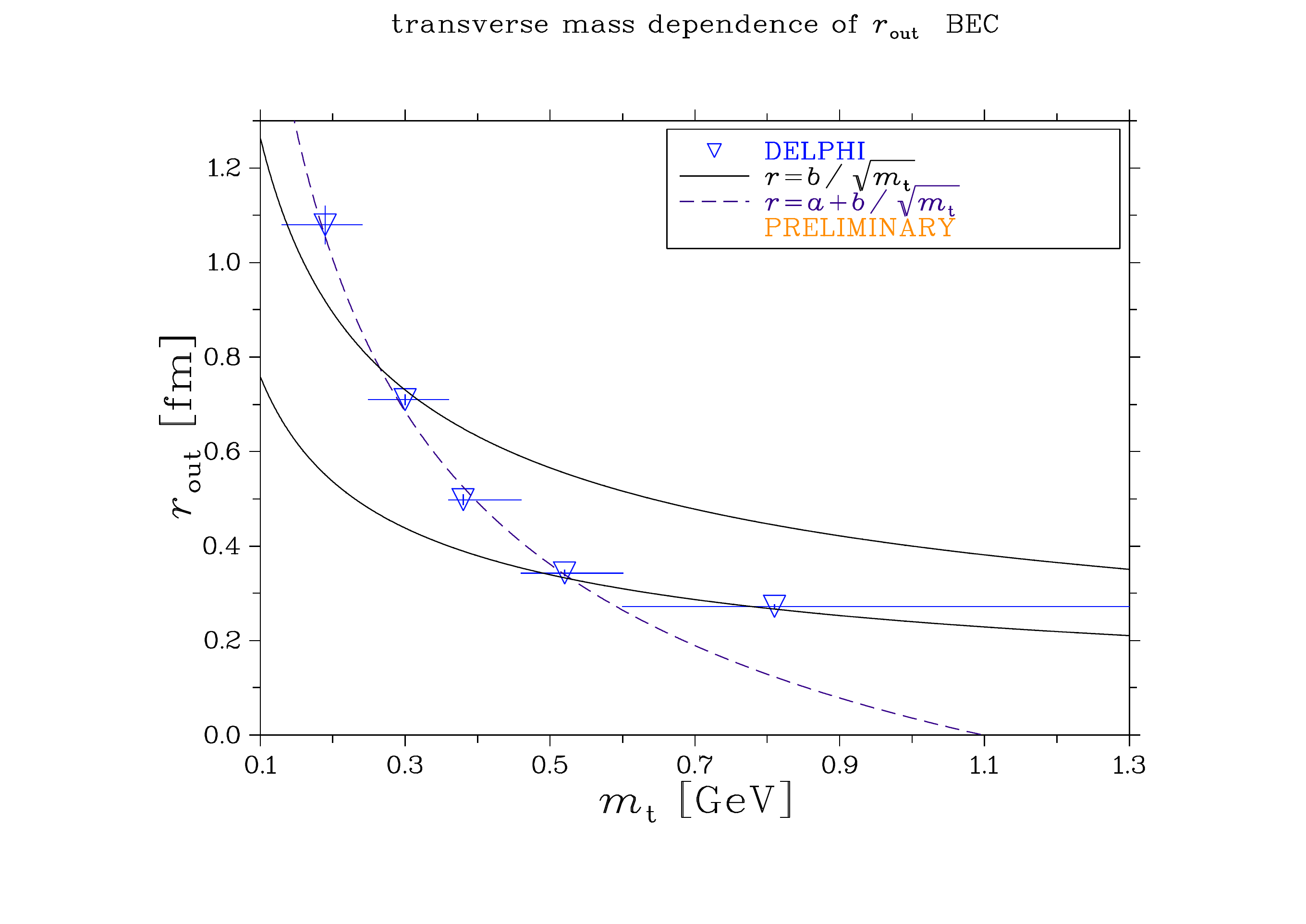}
\end{minipage}
  \caption{The transverse mass dependence of the components of $r$ in the LCMS from Refs.~\citen{Smirnova:Nijm96,Dalen:Maha98}.}
  \label{fig:rmt}
\end{figure*}

\paragraph{Dependence on the transverse mass}
However, $r$ has been observed to depend on the transverse mass of the particle pair,~\cite{Smirnova:Nijm96,Dalen:Maha98}
as is shown in \Fig{fig:rmt}.
 
\paragraph{Dependence on particle and jet multiplicity}
The \OPAL\ collaboration has studied the dependence of $r$ and $\lambda$ on the charged track multiplicity
and on the number of jets~\cite{OPALmult:1996}.
They used an opposite-sign reference sample, which necessitated the exclusion of regions of $Q$ where $R_2$
was too strongly affected by resonances in the reference sample. To describe the long-range correlations they
introduced a quadratic term resulting in
\begin{equation} \label{eq:OPALR2}
   R_2(Q) = \gamma\left[1 + \lambda\exp\left(-(rQ)^2\right)\right]\left(1+\epsilon Q+\delta Q^2\right) \;.
\end{equation}

They observed a linear rise of $r$ with charged track multiplicity
as well as an increase of $r$ with the number of jets.  The behavior of $\lambda$ was the opposite.
However, when only two-jet (or only three-jet) events were selected, $r$ was approximately independent of multiplicity.

\subsection{\taumodel}     \label{taumodel}
However, the ``classic'' parametrization of \Eq{gauss_param} is found to be inadequate, even when it is
generalized to allow for a L\'evy distribution of the source:
\begin{equation} \label{slevy_param}
  R_2 =    \gamma \left[ 1+ \lambda \exp \left(-\left(rQ\right)^{\alpha} \right) \right] (1+ \epsilon Q) \;, \quad 0<\alpha\le2
\end{equation}
This was not realized for a long time because the correlation function was only plotted up to $Q=2\,\GeV$
or less.  In Ref.~\citen{L3_levy:2011} $Q$ was plotted to 4\,\GeV, and it became apparent that there is a
region of anti-correlation ($R_2<1$) extending from about $Q=0.5$ to $1.5\,\GeV$.  This anti-correlation,
as well as the BEC correlation are well described by the \taumodel.

In the \taumodel\ $R_2$ is found to depend not only on $Q$, but also
on quantities $a_1$ and $a_2$.
For two-jet events $a=1/\mt$, where
$\mt=\sqrt{m^2+\pt^2}$ is the transverse mass of a particle).
Parameters of the model are the parameters of the L\'evy distribution which describes the proper time of particle emission:
$\alpha$, the index of stability of the L\'evy distribution; a width parameter $\Delta\tau$; and the proper time $\tau_0$ at which
particle production begins.
 
We shall use a simplified parametrization~\cite{L3_levy:2011} where $\tau_0$ is assumed to be zero and
 $a_1$ and $a_2$ are combined with $\Delta\tau$ to form an effective radius $R$:
\begin{subequations} \label{eq:asymlevR2}
\begin{align}
    R_2(Q) &= \gamma \left[ 1+ \lambda \cos \left(\left(R_\mathrm{a}Q\right)^{2\alpha} \right)
             \exp \left(-\left(RQ\right)^{2\alpha} \right) \right] (1+ \epsilon Q) \;, \label{eq:asymlevR2_}   \\
    R_\mathrm{a}^{2\alpha} &= \tan\left(\frac{\alpha\pi}{2}\right) R^{2\alpha} \;.     \label{eq:asymlevRaR}
\end{align}
\end{subequations}
Note that the difference between the parametrizations of \Eqs{slevy_param} and \ref{eq:asymlevR2}
is the presence of the $\cos$ term, which accounts for the description of the anti-correlation.
The parameter $R$ describes the BEC peak, and $R_\mathrm{a}$ describes the anti-correlation region.
While one might have had the insight to add, \adhoc, a $\cos$ term to \Eq{slevy_param}, it is the \taumodel\
which predicts a relationship,     \Eq{eq:asymlevRaR}, between $R$ and $R_\mathrm{a}$.
 
\begin{figure}     \centering
  \includegraphics*[width=.77\textwidth,clip,viewport=56 87  518 680]{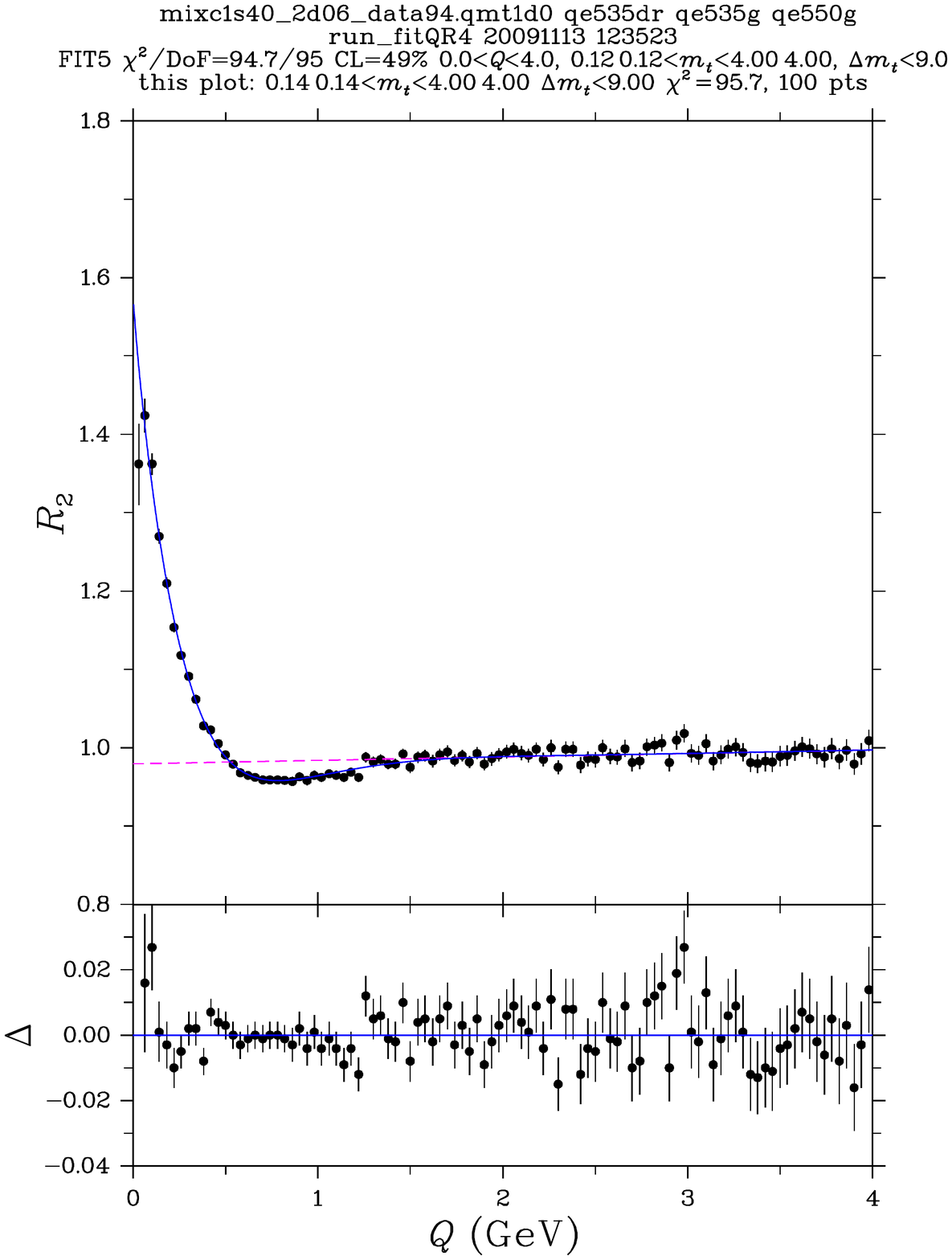}
  \caption{The Bose-Einstein correlation function $R_2$ for two-jet events.
           The curve corresponds to the fit of \Eq{eq:asymlevR2}.
           Also plotted is $\Delta$, the difference between the fit and the data.
           The dashed line represents the long-range part of the fit, \ie, $\gamma(1+\epsilon Q)$.
           The figure is taken from Ref.~\citen{L3_levy:2011}.
          }
  \label{fig:L32jet}
\end{figure}
A fit of \Eq{eq:asymlevR2} to \Lthree\ two-jet events is shown in \Fig{fig:L32jet}, from which
it is seen that the \taumodel\ describes both the BEC peak and the anti-correlation region quite well.
Also the three-jet data is well described~\cite{L3_levy:2011},
which is perhaps surprizing
since the \taumodel\ is inspired by a picture of fragmentation of a single string.
 
It must also be pointed out that the \taumodel\ has its shortcomings:
The \taumodel\ predicts that $R_2$ depends on the two-particle
momentum difference only through $Q$, not through components of $Q$.
However, this is found not to be the case~\cite{L3_levy:2011}.
Nevertheless, regardless of the validity of the \taumodel, \Eq{eq:asymlevR2} provides a good description of the data.
Accordingly, we shall use it in the following.
 
Since the results on the dependence of the BEC parameters on particle and jet multiplicities and on transverse mass
mentioned in Sect.~\ref{review} were obtained using the classic Gaussian parametrization, \Eq{gauss_param},
and since this parametrization has been shown to be inadequate, in the rest of this paper we investigate these
properties using the \taumodel\ parametrization, \Eq{eq:asymlevR2}. The results are \textit{preliminary.}

\subsection{\Lthree\ Data}  \label{l3data}
The data were collected by the
\Lthree\ detector 
at an \Pep\Pem\    center-of-mass energy of $\sqrt{s}\simeq 91.2$\,\GeV\kern-0.1em.
Approximately 36 million like-sign pairs of well-measured charged tracks from about 0.8 million
hadronic Z decays are used. 
This data sample is identical to that of Ref.~\citen{L3_levy:2011}.
 
The same event mixing technique is used to construct $\rho_0$ as in  Ref.~\citen{L3_levy:2011}.

Using the JADE algorithm, events can be classified according to the number of jets.
The number of jets in a particular event depends on the jet resolution parameter of the algorithm, \ycut.
We define \yttJ\ as that value of \ycut\ at which the number of jets in the event changes from two to three.
Small \ytt\ corresponds to narrow two-jet events, large \ytt\ to events with three or more      well-separated jets.
 
\section{New   \textit{Preliminary}\/ Results}                                  \label{NewResults}
\begin{figure}     \centering
  \includegraphics*[width=0.77\textwidth,viewport=-3 0 517 760,clip]{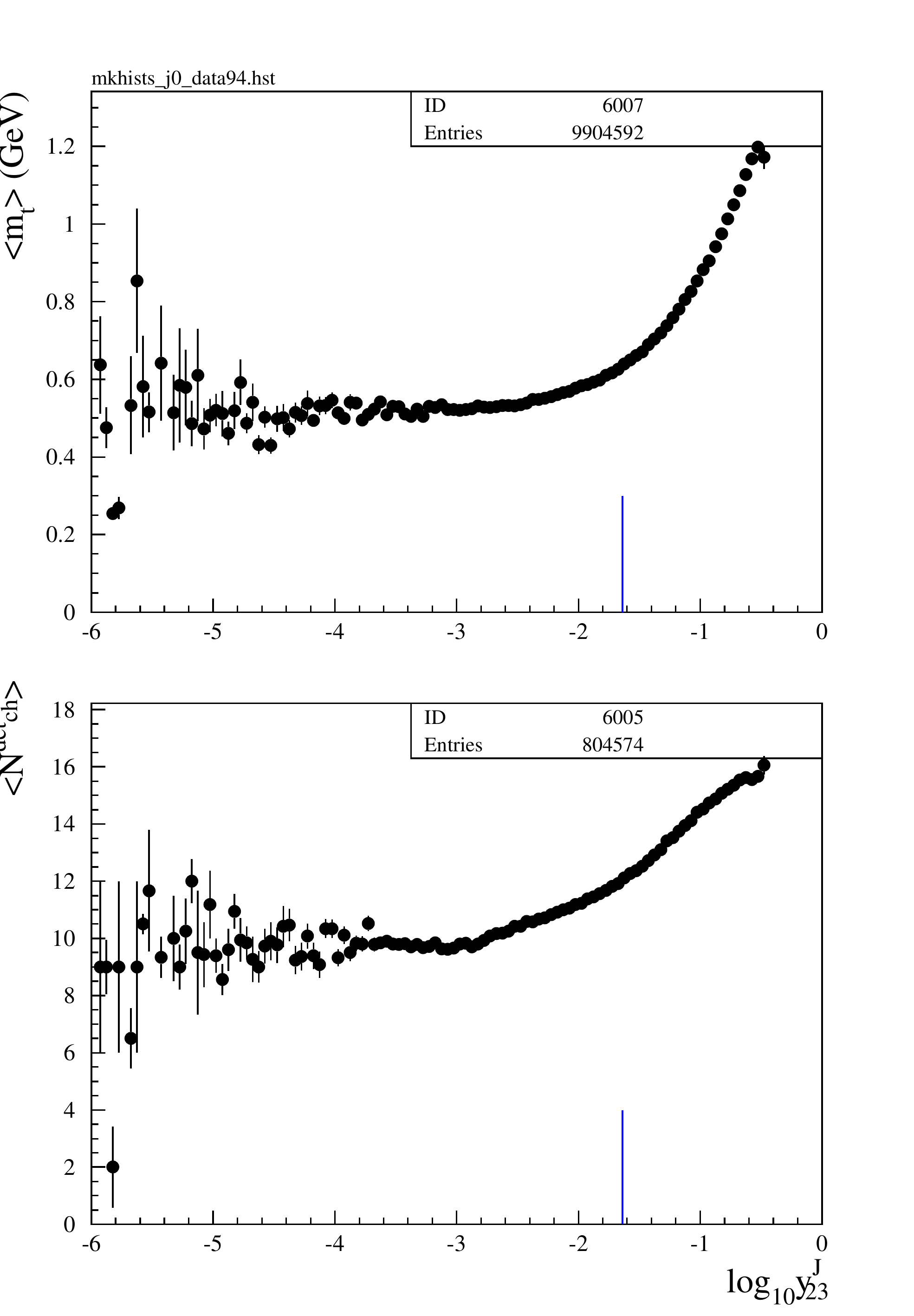}
  \caption{The average transverse mass and
           the average charged multiplicity
           as function of \yttJ.
           The short vertical line at $\log_{10}\yttJ=-1.638$ corresponds to the cut used to define 2- and 3-jet events.
          }
  \label{fig:AvmtNj}
\end{figure}

The parameters of the Bose-Einstein correlation function have been found to depend on
charged multiplicity, the number of jets, and the transverse mass.
However these quantities are related.
Both the charged particle multiplicity and the transverse mass increase rapidly with the
number of jets.
This is seen in \Fig{fig:AvmtNj}, where the average transverse mass and the average charged multiplicity
are plotted \vs\ \yttJ.
In the following we investigate the dependence of $R$ and $\lambda$ on these three quantities.
 
An unfortunate property of the \taumodel\ parameterization, \Eq{eq:asymlevR2}, is that
the estimates of $\alpha$ and $R$ from fits tend to be highly correlated.
Therefore, to stabilize the fits, $\alpha$ is fixed to the value 0.44, which corresponds
to the value obtained in a fit to all events.
 
While we show only the results using the JADE jet algorithm, we have also performed the same analysis using the Durham algorithm.
It is found to lead to the same conclusions.

\subsection{Dependence of \boldmath{$R$} and \boldmath{$\lambda$} on track and jet multiplicities}  \label{depRl}
\begin{figure} \centering
  \includegraphics*[width=.65\textwidth,angle=270,clip,viewport=121 124 555 717]{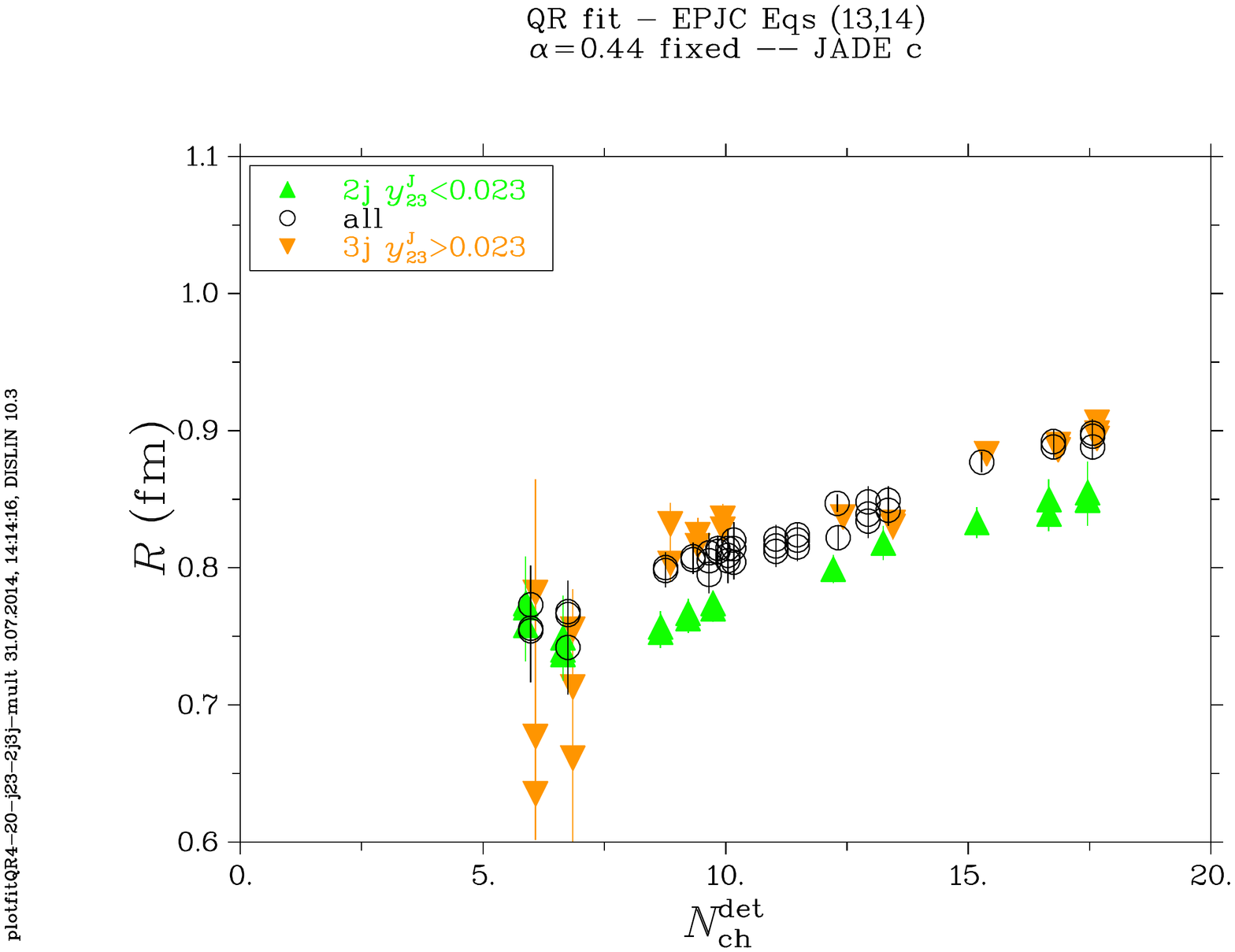}
  \caption{$R$ obtained in          fits of \Eq{eq:asymlevR2}
           as function of detected charged multiplicity
           for two-jet events ($\yttJ<0.23$),
           for three-jet events ($\yttJ>0.23$),  and all events
          }
  \label{fig:RNch}
\end{figure}

\begin{figure} \centering
  \includegraphics*[width=0.65\textwidth,angle=-90,viewport=121 124 555 717,clip]{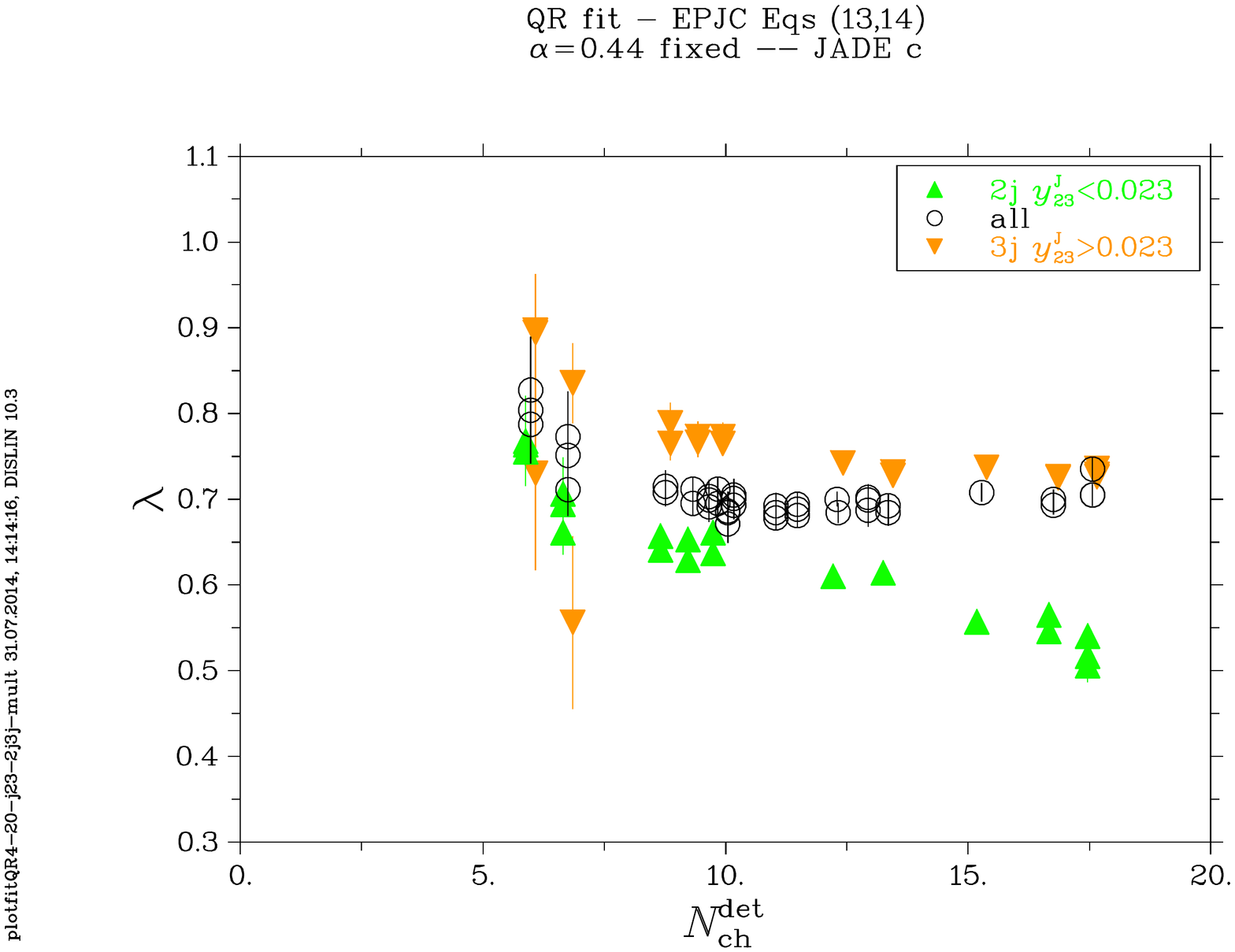}
  \caption{$\lambda$ obtained in          fits of \Eq{eq:asymlevR2}
           as function of detected charged multiplicity
           for two-jet events ($\yttJ<0.23$),
           for three-jet events ($\yttJ>0.23$),  and all events
          }
  \label{fig:LNch}
\end{figure}

The dependence of $R$ and $\lambda$ on the
detected charge multiplicity,\footnote{The charge multiplicity is approximately given by
$N_\text{ch}\approx 1.7 N_\text{ch}^\text{det}$.} is shown in
\Figs{fig:RNch} and~\ref{fig:LNch}, respectively,
for two- and three-jet events as well as for all events.

For all events $R$ is seen to increase linearly with the multiplicity, as was observed for $R$ by \OPAL.
However, the same linear increase is also seen for two- and three-jet events, with $R$ for three-jet events and for all events being
approximately equal and $R$ for two-jet events shifted lower by about a 0.5\,fm.
This contrasts with the \OPAL\ observation of little dependence of $r$ on multiplicity for two- and three-jet events.
 
For all events, as well as for two- and three-jet events, $\lambda$ decreases with multiplicity, the rate of decrease becoming less
for high multiplicity.  It is higher for three-jet events than for two-jet events, with the values for all events lying in between.
This contrasts with the \OPAL\ observation that $\lambda$ was higher for two-jet events, as well as
the \OPAL\ observation that the decrease of $\lambda$ with multiplicity is linear.

\subsection{Dependence of \boldmath{$R$} and \boldmath{$\lambda$} on trasverse mass and jet multiplicity}  \label{depmt}
\begin{figure*}              \centering
\begin{minipage}{.49\textwidth}
  \centerline{\quad JADE 2-jet, $\yttJ<0.023$}
  \vspace{-9mm}
  \includegraphics*[width=0.70\textwidth,angle=-90,viewport=121 123 555 717,clip]{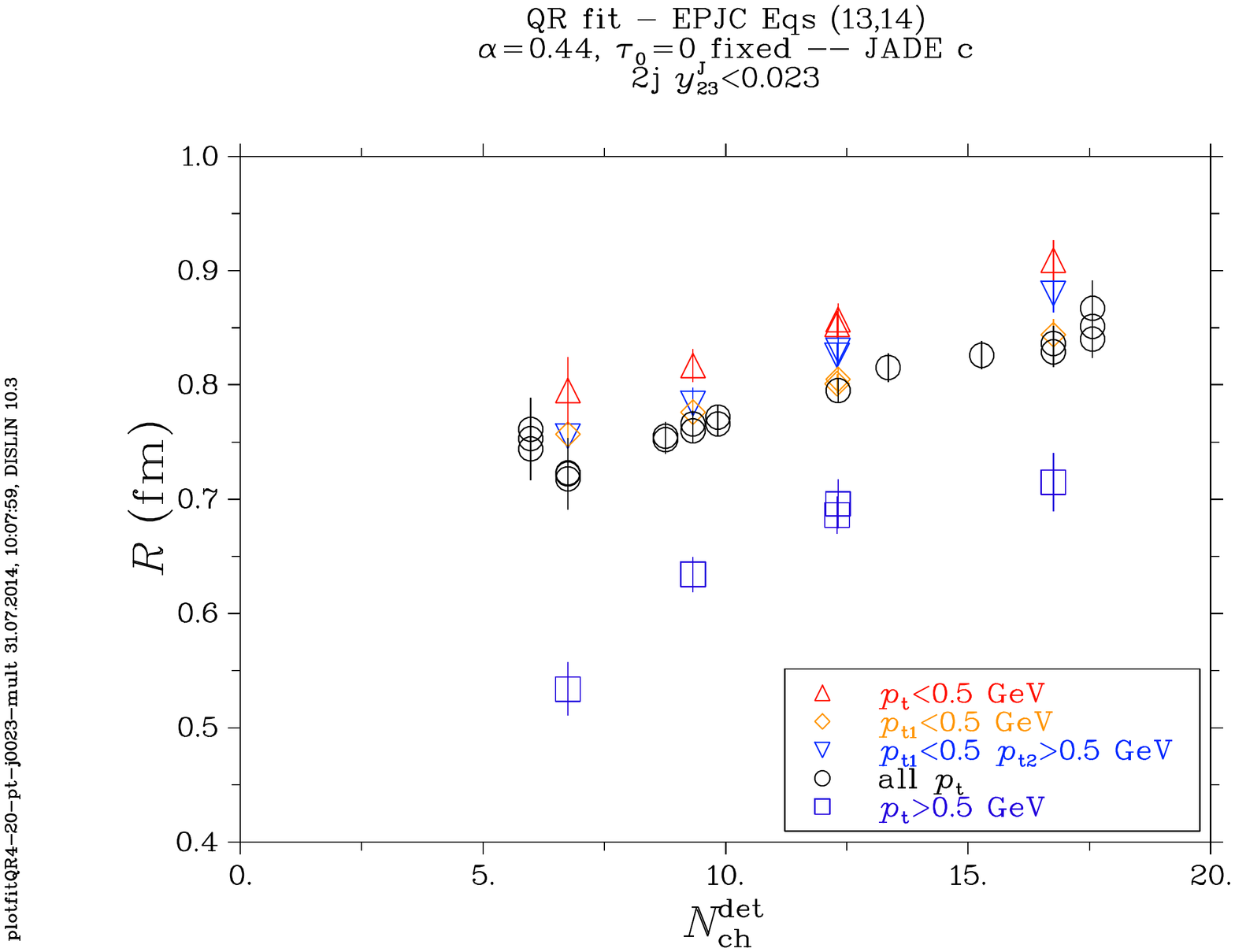}
\end{minipage}
\hfill
\begin{minipage}{.49\textwidth}
  \centerline{\quad JADE 3-jet, $\yttJ>0.023$}
  \vspace{-9mm}
  \includegraphics*[width=0.70\textwidth,angle=-90,viewport=121 123 555 717,clip]{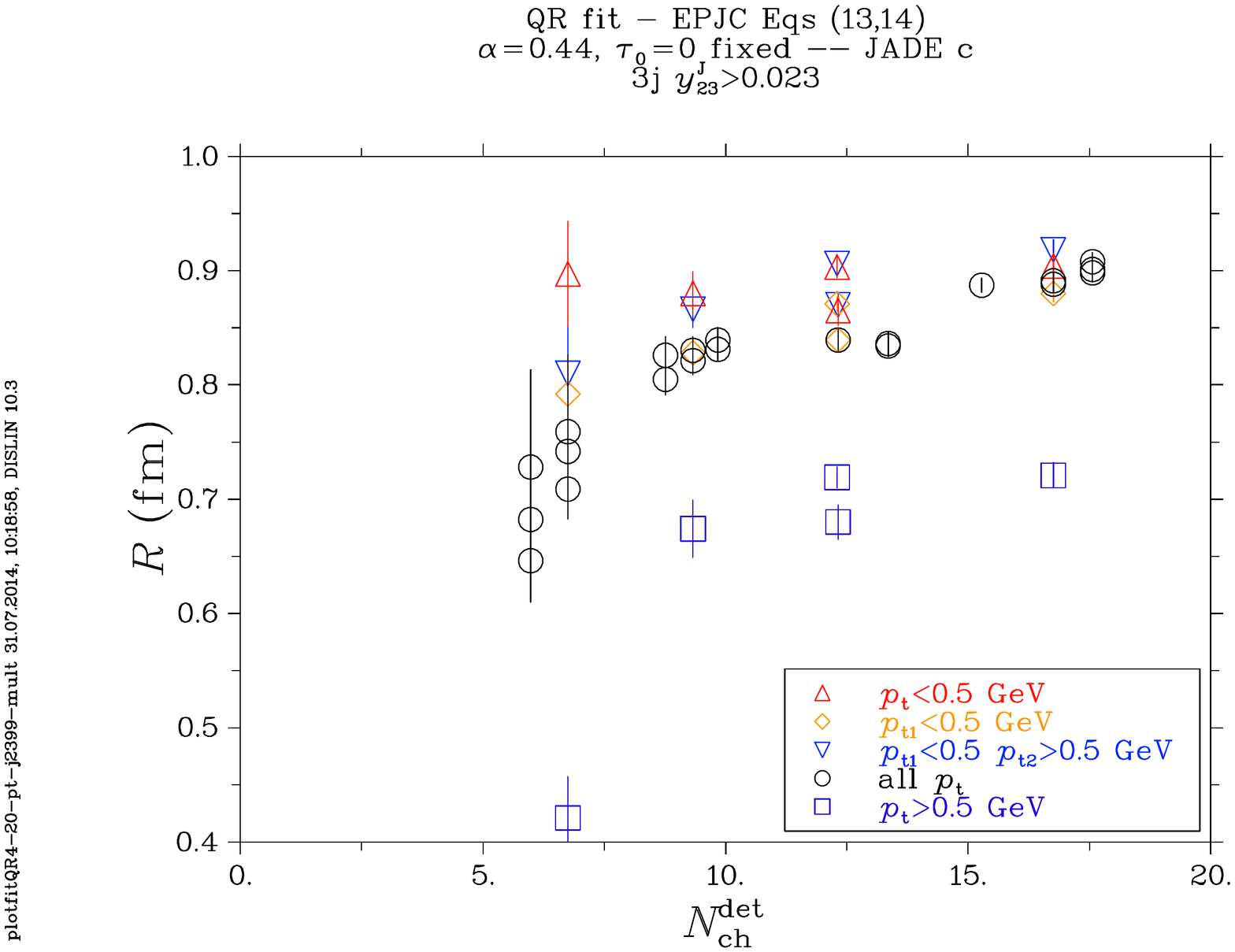}
\end{minipage}
  \caption{$R$ obtained in          fits of \Eq{eq:asymlevR2}
           as function of detected charged multiplicity
           (left) for two-jet events ($\yttJ<0.23$) and
           (right) for three-jet events ($\yttJ>0.23$) with the following selections on \pt:
           $\triangle$ both tracks having $\pt<0.5\;\GeV$;
           $\diamond$ at least one track having
           $\pt<0.5\;\GeV$;
           $\triangledown$ one track with $\pt<0.5\;\GeV$ and one with $\pt>0.5\;\GeV$;
           $\circ$ all tracks;
           $\Box$ both tracks having $\pt>0.5\;\GeV$.
           Note that $\pt=0.5\;\GeV$ corresponds to $\mt=0.52\;\GeV$.
          }
  \label{fig:Rmt}
\end{figure*}

\begin{figure*}              \centering
\begin{minipage}{.49\textwidth}
  \centerline{\quad JADE 2-jet, $\yttJ<0.023$}
  \vspace{-9mm}
  \includegraphics*[width=0.70\textwidth,angle=-90,viewport=121 123 555 717,clip]{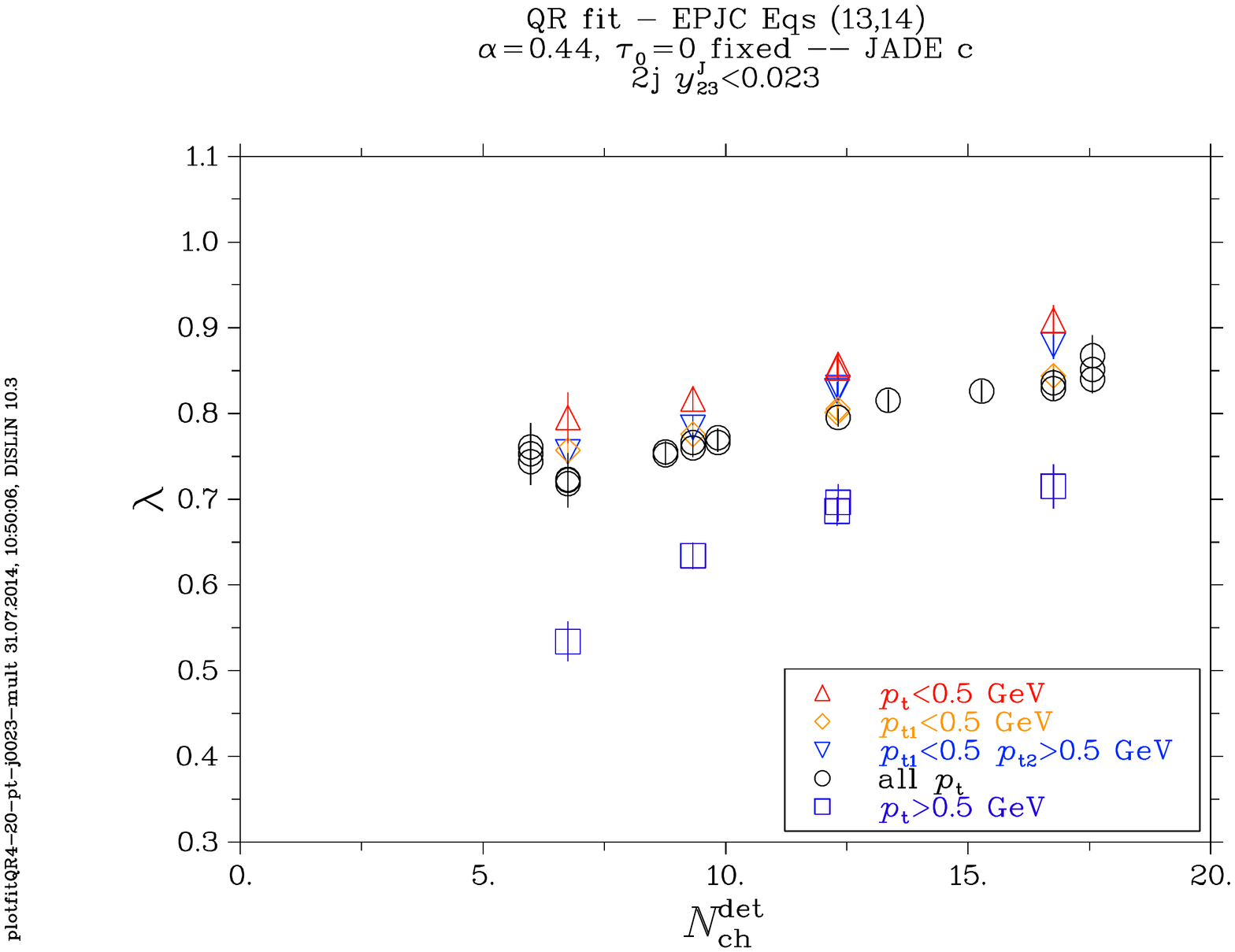}
\end{minipage}
\hfill
\begin{minipage}{.49\textwidth}
  \centerline{\quad JADE 3-jet, $\yttJ>0.023$}
  \vspace{-9mm}
  \includegraphics*[width=0.70\textwidth,angle=-90,viewport=121 123 555 717,clip]{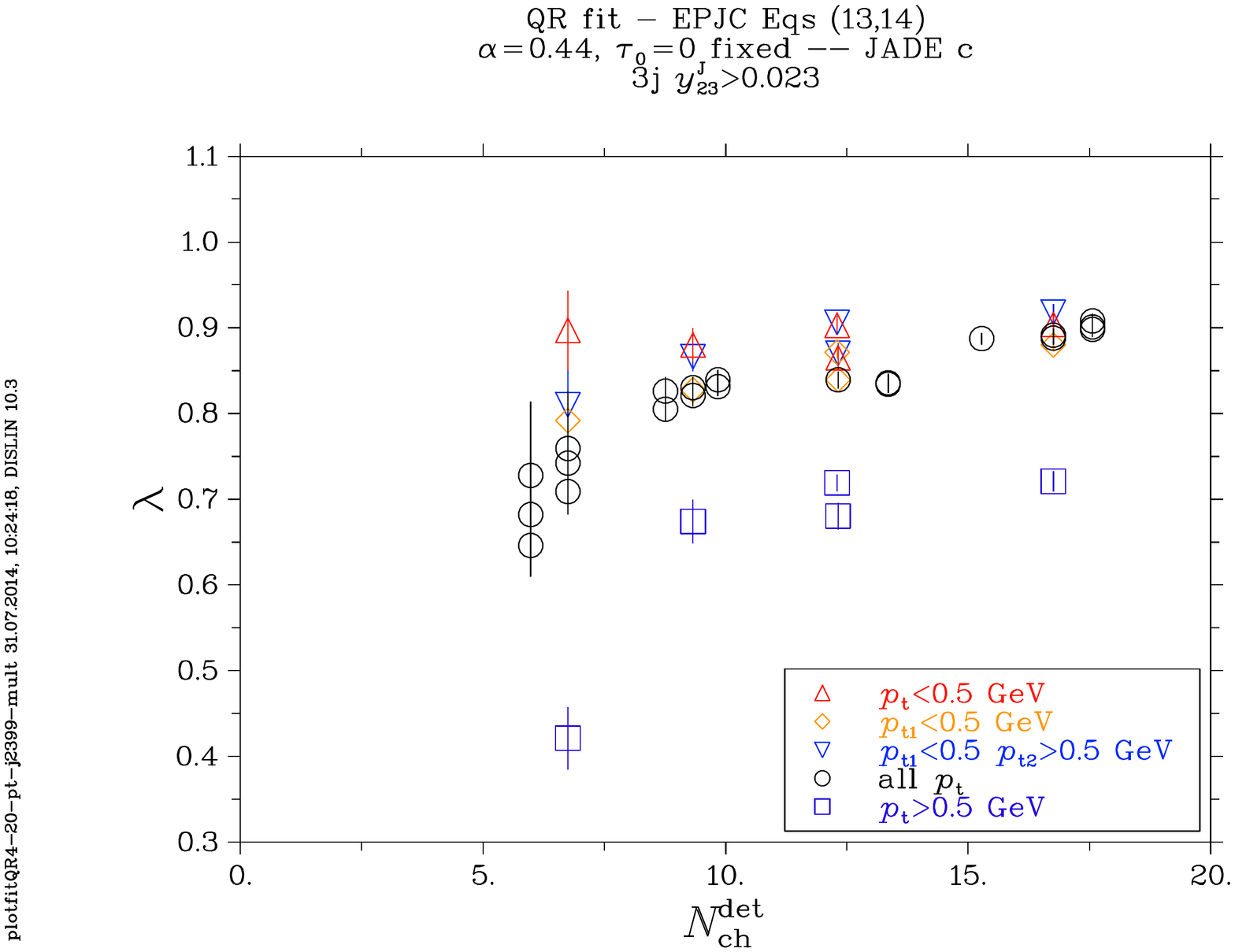}
\end{minipage}
  \caption{$\lambda$ obtained in          fits of \Eq{eq:asymlevR2}
           as function of detected charged multiplicity
           (left) for two-jet events ($\yttJ<0.23$) and
           (right) for three-jet events ($\yttJ>0.23$) with the following selections on \pt:
           $\triangle$ both tracks having $\pt<0.5\;\GeV$;
           $\diamond$ at least one track having
           $\pt<0.5\;\GeV$;
           $\triangledown$ one track with $\pt<0.5\;\GeV$ and one with $\pt>0.5\;\GeV$;
           $\circ$ all tracks;
           $\Box$ both tracks having $\pt>0.5\;\GeV$.
           Note that $\pt=0.5\;\GeV$ corresponds to $\mt=0.52\;\GeV$.
          }
  \label{fig:Lmt}
\end{figure*}

The dependence of $R$ and $\lambda$ on track multiplicity is shown in \Figs{fig:Rmt} and~\ref{fig:Lmt}, respectively,
for various selections on the transverse momentum, \pt, (or, equvilantly, \mt) of the tracks.
For two-jet events both $R$ and $\lambda$ are slightly higher when both tracks have $\pt<0.5\;\GeV$ than when only one track
is required to have so small a \pt.
For three-jet events the same may be true, but the statistical significance is less; the difference decreases with multiplicity.
When neither track has $\pt<0.5\;\GeV$, the values of both $R$ and $\lambda$ are much lower for both two- and three-jet events.
In all cases both $R$ and $\lambda$ increase with multiplicity, and the values for two-jet events are roughly equal to those for
three-jet events.
 
\section{Conclusions}  \label{concl}
The dependence of $R$ and $\lambda$ for the \taumodel\ parametrization is different from that
of $r$ and $\lambda$  found by \OPAL\ for the usual Gaussian parametrization.
However, it is unclear how much the differences depend on the use of different reference samples
and how much on the parametrization used.
 
Multiplicity, number of jets, and transverse mass all affect the values of $R$ and $\lambda$ in the \taumodel\ parametrization,
\Eq{eq:asymlevR2}.
 
\newpage

\bibliographystyle{\lthreebiblio/l3style}
\bibliography{bec,levy}

\end{document}